\documentstyle[LTpaper]{article}
\begin{document}
\input{epsf}
\title{Interplay of charge and spin correlations in nickel perovskites}
\author{J. Loos$^a$ and H. Fehske$^b$}
\address{\mbox{$^a$ Institute of Physics, Czech Academy of Sciences, 16200
Prague, Czech Republic}
\vskip 10pt \mbox{$^b$ Physikalisches Institut, Universit\"at Bayreuth, 
D-95440 Bayreuth, Germany}}
\abstract{
Analyzing the motion of low--spin $(s=1/2)$ holes in a high--spin $(S=1)$
background, we derive a sort of generalized t--J
Hamiltonian for the $\rm NiO_2$ planes of Sr--doped nickelates.
In addition to the rather complex carrier--spin and spin--spin
couplings we take into account the coupling of the 
doped holes to in--plane oxygen breathing modes 
by a Holstein--type interaction term.  
Because of strong magnetic confinement effects the holes are nearly
entirely prelocalized and the electron--phonon coupling becomes much
more effective in forming polarons than in the isostructural
cuprates. In the light of recent experiments on 
$\rm La_{2-x}Sr_xNiO_4$  we discuss how the variety of the 
observed transport and charge/spin--ordering phenomena can be 
qualitatively understood in terms of our model Hamiltonian.
}
\maketitle\pagestyle{empty}
In contrast to the superconducting cuprates, e.g. $\rm
La_{2-x}Sr_xCuO_4$, the layered transition
metal oxide $\rm La_{2-x}Sr_xNiO_4$ becomes metallic only near $x=1$. 
In either case, however, the measurements
of the electrical transport and the magnetic susceptibility, as well as the 
investigations of the lattice and magnetic structures  
revealed a rich variety of phenomena
indicating the close interconnection between the charge-- and
spin--ordering and the transport properties~\cite{Trea95,Chea94}.

The studies on doped $\rm La_{2-x}Sr_xNiO_4$ indicate that 
the additional holes in the $\rm NiO_2$ planes have 
their own magnetic moment, i.e., they carry a
spin $s=1/2$ and couple to the $\rm Ni^{2+}\;(d^8)$ ions 
with spin $S=1$ (the high--spin state, HSS) 
in a way that a low--spin state (LSS) with total spin $J=1/2$ 
is formed~\cite{ZO93}. Excluding the hole doubly--occupied
sites, we shall define our model in the tensor product space of local
hole states $\{|i,\sigma\rangle,\; |i,0\rangle\}$ and local $S=1$
states $\{|i,S,m\rangle\}$. Here, $|i,\sigma\rangle$ means an
eigenfunction of spin operators $\vec{s}_i^{\,2}$, $s_i^z$ of a hole at the
site $i$, $|i,0\rangle$ corresponds to no extra hole at $i$,
and $|i,S,m\rangle$ denotes the eigenfunction of  $\vec{S}_i^2$,
$S_i^z$ of the HSS
with spin projection $m=1,\,0,-1$ at the site $i$.

In the case of nonmaximal total spin $J$, we shall construct our
Hamiltonian in the subspace of the tensor product 
space defined by the local basis vectors 
$\{|i,S,m\rangle|i,0\rangle\}$ and $\{|i,J=1/2,M=\pm
1/2\rangle\}$, i.e., 
\begin{eqnarray}
|i,\mbox{\small $\frac{1}{2}$},M\rangle
\!\!\!\!\!&=&\!\!\!\!\!
\mbox{\small $\frac{1}{\sqrt{3}}$}
\left\{-[\mbox{\small $\frac{3}{2}\!-\!M]^{1/2}$}\,
|i,S,M-\mbox{\small $\frac{1}{2}$}\rangle |i,\uparrow\rangle \right.
\nonumber\\\vspace*{-0.5cm}
&& \quad \;
\left.+ \mbox{\small $[\frac{3}{2}\!+\!M]^{1/2}$}\,
|i,S,M+\mbox{\small $\frac{1}{2}$}\rangle |i,\downarrow\rangle \right\}\!.
\end{eqnarray}
In Ref.~\cite{ZO93}, the hole--transport Hamiltonian given by all processes 
conserving the total--spin $z$--component was expressed by means of 
operators creating HSS from the state of closed Ni--shells
$|0\rangle_S$, namely $B_{im}|0\rangle_S=|i,S,m\rangle$, and by
fermionic operators creating and annihilating the LSS. In contrast
to~\cite{ZO93}, we take into account the `interior structure' of the 
LSS~(1) by the Clebsch--Gordon coefficients. 
This way the hole--transport Hamiltonian can be written
as
\begin{equation}
{\cal H}_t=-t\sum_{\langle i,j\rangle}\left( X^{}_{i\uparrow}
X^{\dagger}_{j\uparrow}+ X^{}_{i\downarrow}
X^{\dagger}_{j\downarrow}\right)
\end{equation}
with 
\begin{eqnarray}
X^{}_{i\uparrow}\!\!\!\!&=&\!\!\!\!
-\mbox{\small $\sqrt{\frac{1}{3}}$}B_{i,0}^\dagger B_{i,0}^{}
h_{i\uparrow}^{} +\mbox{\small $\sqrt{\frac{2}{3}}$}
B_{i,0}^\dagger B_{i,1}^{}
h_{i\downarrow}^{}\nonumber\\
&&\!\!\!\!-\mbox{\small $\sqrt{\frac{2}{3}}$}B_{i,-1}^\dagger B_{i,-1}^{}
h_{i\uparrow}^{}
+\mbox{\small $\sqrt{\frac{1}{3}}$} B_{i,-1}^\dagger B_{i,0}^{}
h_{i\downarrow}^{}\,,
\end{eqnarray}
and the time--reversal transformed expression for $X_{i\downarrow}$,
where the fermionic operators $h_{i\sigma}$ are defined by
$h_{i\sigma}|i,\sigma\rangle=|i,0\rangle$, $h_{i\sigma}^\dagger
|i,0\rangle=|i,\sigma\rangle $.

To discuss the spin dependence of the charge transport 
we use a slightly modified treatment of the spin--charge decoupling proposed
for the t--J model just recently~\cite{Lo96}. In this representation, 
the operators $h_{i\sigma}$ defined above 
are expressed in terms of holon $(e_i)$ and (pseudo) spin--1/2
$(\tilde{s}_i)$ operators 
%\begin{eqnarray}
$h_{i\uparrow}
%\!\!\!\!&=&\!\!\!\! 
=e_i \left(\tilde{s}_i^+ \tilde{s}_i^-
+\mbox{e}^{i\varphi} \tilde{s}_i^- \right)/\sqrt{2}$ and 
%\,,\\
$h_{i\downarrow}
%\!\!\!\!}&=&\!\!\!\! 
=e_i \left(\tilde{s}_i^+ +
\mbox{e}^{i\varphi}\tilde{s}_i^- \tilde{s}_i^+\right)/\sqrt{2}$,
%\,,  
%\end{eqnarray}
where the arbitrary phase factor $\varphi$ does not influence the matrix
elements of the Hamiltonian. Consequently, we
have for the hole spin operators $s_i^+
=h^{\dagger}_{i\uparrow}h_{i\downarrow}^{}=e_i^\dagger
e_i^{}\tilde{s}_i^+$, and the site--occupation operator of LSS is
given by $n_i^{}=e_i^\dagger e_i^{}$. Then the 
total spin operator may be cast into the form
\begin{equation}
\vec{J}_i=(1-n_i)\vec{S}_i+n_i (\vec{S}_i+\vec{\tilde{s}}_i)\,.
\end{equation}  
The spin correlations are determined by antiferromagnetic (AF)
exchange interactions
\begin{equation}
{\cal H}_{ex}=\sum_{\langle i,j \rangle} {\cal J}(n_i,n_j)
\vec{J}_i\vec{J}_j\,,
\end{equation} 
where the operators $\vec{J}_i$ are given by~(4) and the arguments
$n_i$, $n_j$ of ${\cal J}$ indicate the dependence of the exchange on the
electronic configuration of the nearest neighbour (NN) sites $\langle
i,j\rangle$. We have to keep in mind that $\vec{s}_i$, $\vec{S}_i$
couple one another to form a $J=1/2$ state, what can be enforced 
by adding an effective on--site interaction ${\cal J}_i n_i
\vec{\tilde{s}}_i \vec{S}_i$ which is much stronger than all the
inter--site interactions.

The influence of the magnetic correlations on the charge transport will be
demonstrated for low doping $x\ll 1$ in which case the holes are
moving in the AF background of $\rm Ni^{2+}$ spins. Using the linear
spin wave approximation (LSWA) and the representation of
$h_{i\sigma}$, $h_{j\sigma}$ in terms of decoupled spin-- and charge
variables, ${\cal H}_t$ given by (2), (3)  assumes 
the form of a spin--dependent transport Hamiltonian for holons. The
effective charge--transfer constant is obtained by taking the average of
${\cal H}_t$ over the spin degrees of freedom. The average over the
spin $S=1$ background in LSWA leads to
\begin{equation}
\!{\cal H}_t= -\frac{\sqrt{2}}{3} t \sum_{\langle i,j\rangle} 
e_i^{}e_j^\dagger (\langle \delta S_i^z \rangle + \mbox{\small
$\frac{1}{2}$} \langle S_i^+ S_j^-
\rangle)(f_{ji}+g_{ji}),
\end{equation}
where
\begin{eqnarray}
f_{ji}\!\!\!\!&=&\!\!\!\!\vec{\tilde{s}}_i\vec{\tilde{s}}_j +
\frac{1}{4}\,,\\[0.1cm] 
g_{ji}\!\!\!\!&=&\!\!\!\!\frac{\mbox{e}^{i\varphi}}{2}
\left[\left(\frac{1}{2}-\tilde{s}_i^z\right)
\tilde{s}_j^-+\tilde{s}_i^-\left(\frac{1}{2}+\tilde{s}_j^z\right)\right]
\nonumber\\
&&\!\!\!\frac{\mbox{e}^{-i\varphi}}{2}
\left[\left(\frac{1}{2}+\tilde{s}_i^z\right)
\tilde{s}_j^++\tilde{s}_i^+\left(\frac{1}{2}-\tilde{s}_j^z\right)\right]\!,
\end{eqnarray}
and $\langle \delta S_i^z \rangle$ is equal to the reduction of the
local $|S_i^z|$ from the classical value $S$ in the AF magnon ground
state. Both the expectation values in (6) 
(being zero in the classical AF N\'{e}el ground
state) are given by the zero--point spin
fluctuations and cause a reduction of the bare hopping constant~$t$. A
similar effect is induced also by the remaining spin factor, as the
mean value
$
\langle f_{ji}+g_{ji} \rangle = \langle 2 \vec{\tilde{s}}_i  
\vec{\tilde{s}}_j +\mbox{\small $\frac{1}{2}$} \rangle
$
and the spins $ \vec{\tilde{s}}_i,\,  \vec{\tilde{s}}_j$ ought to be
AF correlated owing to their coupling to the $S=1$ spins.

The latter arguments concerning the reduction of the hole hopping rate
may be qualitatively also applied to the AF long--range order (LRO)
for $x\to1$, to the ferrimagnetic LRO at $x=1/2$ as well as to the
more general commensurate LRO 
(for $x=1/3$) or incommensurate ordering of LSS and HSS with a large
correlation length, provided that the spins are mostly AF correlated.

Electron diffraction studies and x--ray scattering technique
revealed~\cite{Isea94} that the observed variety of magnetic
structures in the nickelates is closely related 
to the lattice structure modulation which
indicates the ordering of local lattice deformations connected with
the formation of quasi--localized polarons. These findings may be
understood in the light of the above model considerations if the
interaction of holes with the lattice is taken into account.

In fact, the spin correlations suppressing the charge transport
facilitate the polaron self--trapping by electron--phonon
coupling~\cite{Feea95}. 
The Holstein--type interaction of holes with an in--plane
(breathing) mode has the form
\begin{equation}
{\cal H}_{h-p}=-\sum_{\langle i,j\rangle} A u_{ij}
(e_i^+e_i^{}-e_j^+e_j^{})\,,
\end{equation}
where $u_{ij}$ means the displacement of the oxygen in the bond between
NN ions $i,\,j$. Consequently, the deformation of the bonds leading to
the hole localization and energy lowering is given by the charge
difference between the NN sites. Thus we expect  the polaron ordering
given by the energy optimization of the distribution of HSS and
LSS. Certain analogy with the usual phase separation exists: in the
latter case the optimization of bond distribution is given by the
exchange energy, while for the nickelates, the energy gain connected
with the localized polaron formation seems to be most important.

\end{document}